\documentclass[10pt,conference]{IEEEtran}
\usepackage{fontawesome}
\usepackage{geometry}
\usepackage{pifont}
\usepackage{amssymb}
\usepackage{comment}
\usepackage{caption}
\IEEEoverridecommandlockouts
\usepackage{cite}
\usepackage[utf8]{inputenc}
\usepackage{amsmath,amssymb,amsfonts}
\usepackage{algorithmic}
\usepackage{graphicx}
\usepackage{textcomp}
\usepackage{subcaption}
\usepackage{xcolor}
\usepackage{titleps}
\usepackage{multirow}
\usepackage{booktabs}
\usepackage{array}
\usepackage{etoolbox}
\usepackage{dblfloatfix}
\usepackage{floatpag}
\usepackage{float}

\usepackage[
    colorlinks,
    linkcolor=black,   
    citecolor=green,   
    urlcolor=black,    
    filecolor=black,   
    menucolor=black,   
    runcolor=black     
 ]{hyperref}
\begin{document}

\title{Wav2DF-TSL: Two-stage Learning with Efficient Pre-training and Hierarchical Experts Fusion for Robust Audio Deepfake Detection}


\makeatletter

\makeatother

\author{
    \IEEEauthorblockN{
        Yunqi Hao\IEEEauthorrefmark{1,2}$^{*}$\thanks{$^{*}$Equal Contribution},
        Yihao Chen\IEEEauthorrefmark{2,3}$^{*}$,
        Minqiang Xu\IEEEauthorrefmark{2}\textsuperscript{\dag}\thanks{\textsuperscript{\dag}Corresponding Author}, 
        Jianbo Zhan\IEEEauthorrefmark{2}\textsuperscript{\dag},
        Liang He\IEEEauthorrefmark{1,4}, 
        Lei Fang\IEEEauthorrefmark{2}, 
        Sian Fang\IEEEauthorrefmark{2},
        Lin Liu\IEEEauthorrefmark{2}
    }
     \IEEEauthorblockA{}  \IEEEauthorblockA{\IEEEauthorrefmark{1}School of Computer Science and Technology, Xinjiang University, Urumqi, China}
    \IEEEauthorblockA{\IEEEauthorrefmark{2}Hefei iFly Digital Technology Co. Ltd., Hefei, China}
    \IEEEauthorblockA{\IEEEauthorrefmark{3}University of Science and Technology of China, Hefei, China}
    \IEEEauthorblockA{\IEEEauthorrefmark{4}Department of Electronic Engineering, Tsinghua University, Beijing, China}}

\maketitle

\begin{abstract}
In recent years, self-supervised learning (SSL) models have made significant progress in audio deepfake detection (ADD) tasks. However, existing SSL models mainly rely on large-scale real speech for pre-training and lack the learning of spoofed samples, which leads to susceptibility to domain bias during the fine-tuning process of the ADD task. To this end, we propose a two-stage learning strategy (Wav2DF-TSL) based on pre-training and hierarchical expert fusion for robust audio deepfake detection. In the pre-training stage, we use adapters to efficiently learn artifacts from 3000 hours of unlabelled spoofed speech, improving the adaptability of front-end features while mitigating catastrophic forgetting. In the fine-tuning stage, we propose the hierarchical adaptive mixture of experts (HA-MoE) method to dynamically fuse multi-level spoofing cues through multi-expert collaboration with gated routing. Experimental results show that the proposed method significantly outperforms the baseline system on all four benchmark datasets, especially on the cross-domain In-the-wild dataset, achieving a 27.5\% relative improvement in equal error rate (EER), outperforming the existing state-of-the-art systems.

\end{abstract}
\section{Introduction}

\label{sec:intro}
Audio deepfake detection (ADD) is a technique that uses artificial intelligence algorithms to analyze speech signals to determine whether they have been spoofed or tampered with. Early speech synthesis techniques faced on intelligibility and naturalness issues and were easily detected. However, with the development of deep learning in text-to-speech (TTS) and voice conversion (VC), synthesized speech has achieved significant improvements in naturalness and fluency. To address the challenge that realistic synthetic speech not only deceives human auditory perception, but also may attack automatic speaker verification (ASV) systems. The ASVSpoof community has organized a series of anti-spoofing challenges\cite{19LA,21LA,asvspoof5}, created public datasets and evaluation standards, promoted research and development on ADD task.

Common methodes for ADD systems can be categorized into three types. The first type relies on hand-crafted features, including linear frequency cepstral coefficients (LFCC) \cite{LFCC_s2pecnet,LFCC_2}, constant-Q transform (CQT)\cite{CQT}, and short-time fourier transform (STFT)\cite{sTFT_fastaudio,sTFT_2}. The second type is based on end-to-end methods, such as utilizing SincNet\cite{sincNet_1,sincNet_AASIST,sincNet_DFsincNet}. However, the performance of both types degrades rapidly when exposed to unseen spoofing attacks or disturbances caused by encoding and transmission. The third type is based on features extracted from pre-trained speech models. In recent years, self-supervised learning (SSL) has achieved significant advancements in fields such as automatic speech recognition (ASR) and speaker verification (SV)\cite{ssL_AsR,ssL_sV}. such as Wav2vec2.0\cite{Wav2vec2.0}, HuBERT\cite{hubert} and WavLM\cite{WavLM} have exhibited promising performance in various speech processing tasks. It has been demonstrated that the application of pre-trained models also has been shown to enhance the detection performance against ADD systems to some extent\cite{w2v+MFA,XLSR+AASIST,XLSR+LLGF}.\par

Recent research shows that using speech pre-training models as front-end feature extractors can improve the performance of ADD systems. However, existing SSL models mainly rely on large-scale real speech for pre-training and lack the learning of spoofed samples in specific scenarios, resulting in models that are prone to domain bias, especially spoofed speech that has been processed by codecs and compression. Our goal is to incorporate spoofed samples to mitigate the domain discrepancy. However, continuous training directly on the original model is not only computationally expensive, but also may trigger catastrophic forgetting. In recent years, in the fields of natural language processing (NLP) and computer vision (CV), parameter efficient fine-tuning (PEFT)\cite{LoRA, Adapter} has been demonstrated to be able to achieve comparable performance to full fine-tuning in multiple downstream tasks, and freezes most of the parameters to effectively mitigate the above problem. Inspired by this, we introduce LoRA and adapter fine-tuning in the self-supervised learning stage of Wav2vec2.0 to efficiently learn artefacts of spoofed speech while preserving the original pre-training knowledge for robust audio deepfake detection.\par

It has been shown that the hidden embeddings extracted by self-supervised pre-training models are able to capture multi-level information in speech, such as phonemes, semantics, emotions and speaker attributes. Utilizing different levels of hidden layer features is beneficial for ADD tasks\cite{XLSR+SLS,wavlm+layershidden}. However, current fusion methods lack effective selection of features and tend to introduce too much redundant information leading to poor performance. To cope with this challenge, the mixture of experts (MoE)\cite{MoE} is emerging as an effective solution. MoE is a widely used integration method, which is usually regarded as an ensemble of multiple sub-networks (experts) and the weights are assigned to these experts through a trainable gated network to enhance the performance of a specific task during model training.\par
In this paper, we propose a two-stage learning strategy based on the XLSR pre-trained model, named Wav2DF-TSL, to enhance its performance in the field of deepfake speech detection. The strategy consists of two phases: self-supervised learning and supervised fine-tuning. In the self-supervised learning stage, we leverage a large amount of unlabeled spoofed samples as input and incorporate LoRA and convolutional adapter  to efficiently capture the local and global dependencies of spoofed speech. This method optimizes the SSL model's adaptability to artifacts while effectively avoiding catastrophic forgetting. In the fine-tuning stage, we design a hierarchical adaptive mixture-of-experts method, which uses a gating network assisted by hierarchical contributions to dynamically select the most suitable experts for fusing multi-layer hidden features, mitigating the interference of redundant information. In summary, the main contributions of this paper are as follows:
\begin{itemize}
\item[$\bullet$] We proposed a two-stage learning strategy (Wav2DF-TSL) combining self-supervised learning and fine-tuning for audio deepfake detection. This strategy effectively exploits artifact information from both unlabeled and labeled spoofed samples, enhancing the model's generalization ability and robustness.
\item[$\bullet$] The proposed an efficient self-supervised learning method based on the XLSR model, which effectively learns the artifacts of spoofed samples by inserting adapters while mitigating catastrophic forgetting. Experiments show that adding unlabeled spoofed data can enhance SSL front-end features for ADD tasks.
\item[$\bullet$] The proposed a hierarchical adaptive mixture-of-experts (HA-MoE) method. This method optimizes the attention of hierarchical features in the SSL model for the ADD task and combines them with the gating network, assisting the dynamic routing selection of the expert model, enhancing the model's adaptability to spoofed features.
\item[$\bullet$] Experiments demonstrate that the proposed Wav2DF-TSL method achieves consistent performance improvements across four benchmark datasets (ASVspoof 19LA, 21LA, 21DF, and In-the-wild). It achieves an average relative improvement of 27.8\% in EER compared to the baseline system. Even with only 12\% of the parameters, it achieves a 19.5\% relative improvement in average EER and outperforms other advanced systems.

\end{itemize}
\begin{figure*}[!t]
    \clearpage
    \centering
    \includegraphics[width=0.74\textwidth, height=0.33\textheight]{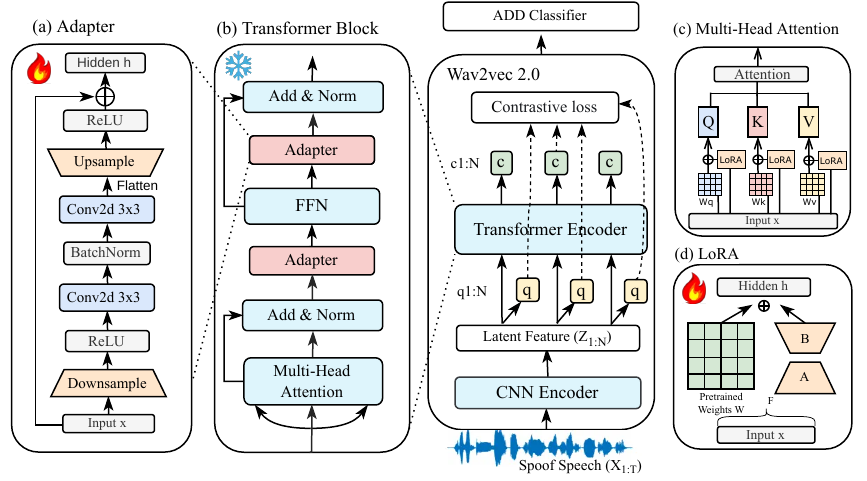} 
    \caption{The framework of parameter-efficient self-supervised pretraining. (a) and (b) represent the in-block adapter network and its embedding position in each transformer layer, while (c) and (d) represent the LoRA structure and its embedding position in the multi-head attention layer. This stage optimizes self-supervised representations from large-scale unlabeled spoofed samples, helping the ADD classifier extract more discriminative spoofing features.}
    \label{fig:Architecture}
\end{figure*}

\section{Proposed Method}
\label{sec:format}

\subsection{Wav2vec2.0 Model}
Wav2vec 2.0 is a self-supervised learning framework for speech representation developed by Facebook, primarily used for speech recognition tasks. This method extracts general speech representations from a large amount of unlabeled raw audio data, effectively enhancing speech feature extraction capabilities while significantly reducing reliance on large-scale labeled data. The model consists of two components: a convolutional encoder and a context encoder. The convolutional encoder extracts a sequence of feature vectors \( Z_{1:N} \) from the input waveform \( X_{1:T} \), while the context encoder transforms \( Z_{1:N} \) into the output \( C_{1:N} \), capturing information from the entire sequence. The ratio between \( N \) and \( T \) is determined by the stride of the convolutional network, with the default setting being \( N/T = 1/320 \). During training, the model masks part of \( Z_{1:N} \), and then computes a new sequence \( C_{1:N} \) using a transformer based on the partially masked \( Z_{1:N} \). Additionally, the model quantizes the latent vectors \( Z_{1:N} \) into \( Q_{1:N} \), and uses a contrastive loss to evaluate how well the model can identify each \( Q_n \) among multiple distractors given \( C_n \).

\par
The goal of this paper is to extend self-supervised representation learning to the ADD task. We selected the XLSR series model as the initial weights for self-supervised pretraining. This model is pre-trained on 436k hours of unlabeled speech data across 128 languages. Compared to other speech pre-training models, XLSR expands the number of languages, the scale of training data, and the model parameters, thus offering stronger domain generalization and robustness.

\subsection{Parameter-Efficient Self-supervised Learning}
To address the limitation of current SSL models in representing deepfake samples in specific scenarios, such as speech deepfake detection, this study proposes a self-supervised pretraining strategy based on parameter-efficient transfer learning, as shown in Figure~\ref{fig:Architecture}. The method introduces a large amount of task-relevant, unlabeled spoof speech as input and continues training the original XLSR model weights. To maximize the preservation of useful information in the pretrained model, an adapter network is integrated into the transformer layers of the XLSR model. During training, only the adapter parameters are updated, enabling lightweight and efficient feature optimization. Specifically, this paper investigates two parameter-efficient transfer learning methods: LoRA and Adapter.\par
\textbf{Low-rank Adaptation (LoRA)}. In the self-supervised pretraining stage, we combine the LoRA module with the context encoder of the XLSR pretrained model, using a learnable low-rank matrix to effectively learn task-related speech spoofed features. The LoRA structure is inserted in parallel into the multi-head attention (MHA) layers of the transformer encoder, applied separately to the query, key, and value vectors. The hidden embeddings are adjusted by updating the low-rank matrix, while keeping the original pretrained model weights unchanged, to enable domain-specific adaptation.\par
Specifically, we use low-rank decomposition to effectively limit the model's update range for the pre-trained weights, allowing the model to focus on patterns related to forged data while avoiding overfitting to the noise in the training data. Suppose the given hidden layer weight matrix is \( W_0 \). LoRA introduces two low-rank matrices \( A \) and \( B \), where matrix \( A \in \mathbb{R}^{r \times d_{in}} \) and \( B \in \mathbb{R}^{d_{out} \times r} \), \( r \ll \min(d_{in}, d_{out}) \). Low-rank matrices A and B are trainable while the original
weight \( W_0 \) and bias \( b \) remain unchanged during training.  The specific forward computation process is as follows:
\[h = W_0 x + \Delta W x + b = W_0 x + B A x + b\tag{1}\]
\(\Delta W = B A\) represents the update part through the low-rank matrices. This design allows the model to significantly reduce computational resources and effectively focus on the discriminative spoofed information.\par

\begin{figure*}[!t]
    \clearpage
    \centering
    \includegraphics[width=0.74\textwidth, height=0.34\textheight]{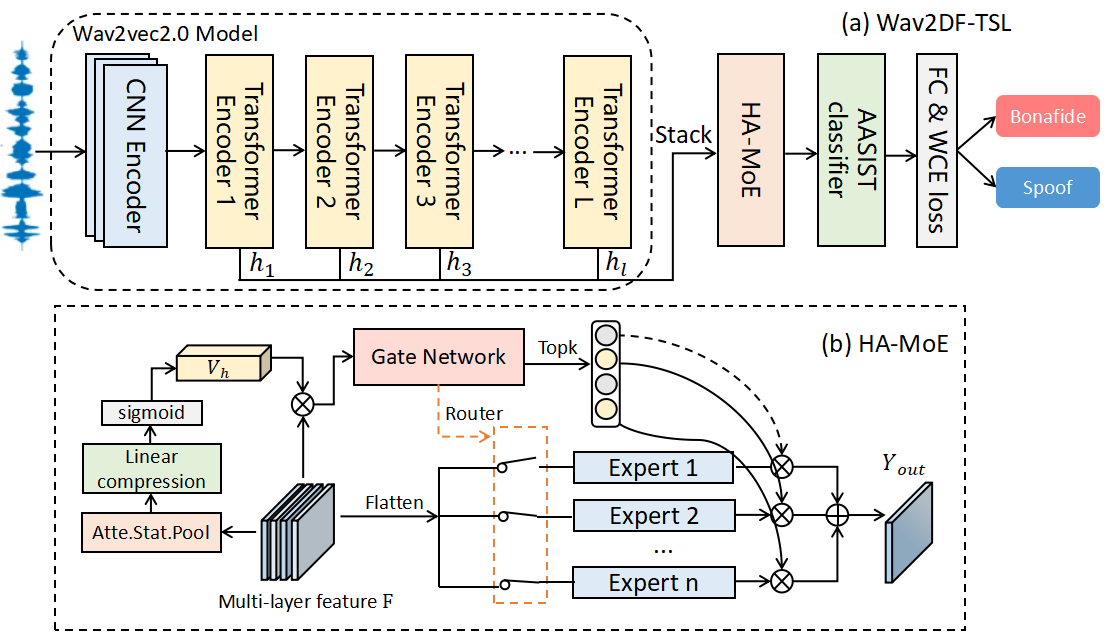} 
    \caption{Overall architecture of the proposed ADD system. (a) represents the pipline of the fine-tuning phase of Wav2DF-TSL. (b) represents the implementation process of the HA-MoE method, which is used to process the stacked multi-layer hidden embeddings, fusing the hierarchical artifacts of the spoofed samples.}
    \label{fig:HA-MoE}
\end{figure*}

\textbf{Adapter Tuning}.
spoofed speech samples often contain subtle local artifacts and are easily influenced by codecs and noise. Directly applying adapter fine-tuning methods from NLP to the ADD task may struggle to effectively capture the multi-level spoofing cues present in speech signals. To address this issue, we designed a bottleneck convolutional adapter that focuses on capturing both local and global dependencies in spoofed speech. During training, we freeze the rest of the model and only update the weights of the adapter within the block. This method allows the SSL model to retain more pre-trained knowledge, similar to a repository for speech spoofing features, enabling adaptation to specific domains.\par
The adapter consists of a downsampling linear layer (Down), an upsampling linear layer (Up), 2D-convolution layers, ReLU activations, and batch normalization. This structure is inserted after the multi-head attention layer (MHA) and feed-forward layer (FFN) in each transformer block of the XLSR model. The convolutional encoded features, after being masked, are fed into the contextual network to generate hidden embeddings \( x \), \( x \in \mathbb{R}^{T \times F} \). The embeddings \( x \) are compressed to dimension \( s \) by a downsampling layer, lowering computational costs while extracting key information. After channel expansion, the features are processed by two \( 3 \times 3 \) 2D-convolutional layers to model local dependencies in the time-frequency dimensions. The features are then flattened along the channel dimension and restored to their original size using an upsampling layer. Finally, a residual connection preserves the global knowledge of the pre-trained model while improving the ability to capture spoofing features. The detailed forward process is as follows:
\[x' = BN(Conv2d(\sigma(x\cdot W_{Down})))\tag{2}\]
\[h = \sigma(\psi(Conv2d(x'))\cdot W_{Up})+x\tag{3}\]
Here, \( \sigma \) represents the ReLU activations. \( \psi \) represents the flatten operation.\par
Finally, the model uses context features \(c_t \) to predict the masked positions of the quantized features \( q_t \) and computes contrastive loss, thereby enhancing the model's ability to represent multi-level spoofing features and noise robustness. The contrastive loss formula is as follows:

\[L_c = - \log{\frac{\exp(sim(c_t,q_t)/ k) } {\sum_{q\in Q_t}^{} \exp(sim(c_t,\tilde{q} )/ k)} }\tag{4}\]

Here, $k$ is constant temperature during training, and \(sim \) denotes the cosine similarity calculation. $\tilde{q}$ is the correctly predicted quantized representation $q_t$ from the candidate quantized representations.

\subsection{Hierarchical Hidden Embeddings of SSL Model}
Recent research has shown that hidden embeddings extracted by self-supervised pre-trained models can capture speech information at various levels, such as phonetic, semantic, emotional, and speaker attributes. These hierarchical features can enhance the feature representation ability for different downstream tasks. However, during the fine-tuning of the SSL model, current methods do not fully consider the contribution of hidden embeddings from different levels to the ADD task. Our goal is to design a learnable hierarchical contribution vector \( V_{h} \) to maximize task-relevant discriminative features, enhancing the expression of deepfake features.
\par
As shown in the Figure~\ref{fig:HA-MoE}(a), The process takes the raw waveform as input and is processed through multiple stacked transformer layers within the SSL model to extract contextual sequence information. The hidden embedding tensor at the $l$-th layer of the transformer is denoted as $h_l \in \mathbb{R}^{T \times D}$, where $l = \{1, 2, \dots, L\}$. $L$ represents the number of transformer layers, and $D$ is the hidden dimension. For the XLSR pre-trained model, $L = 24$ and $D = 1024$. Next, the stacked hidden embeddings \(F \in \mathbb{R}^{L \times T \times D}\) will be processed through a two-step compression process.\par
The stacked hidden embeddings are processed using adaptive statistical pooling (ASP) along the time dimension $T$. ASP dynamically adjusts the size of the pooling region, allowing it to compress the time dimension based on the distribution of the input features, enabling the extraction of more discriminative spatiotemporal features for different speech patterns. Then, the hidden dimension $D$ is further compressed using a fully connected layer, obtaining $V_l$, where $V_l \in \mathbb{R}^{L \times 1 \times 1}$. The specific formula is as follows:
\[V_l = \left( \text{ASP}_T \left( \sum_{l=1}^{L} \left( h_l \cdot W_l \right) \right) \right) \cdot W_s
\tag{5}
\]
Here, \( W_l \) and \( W_s \) represents the weight of the compression linear layer, \( W_l \in \mathbb{R}^{D \times D_{h}} \). \( W_s \in \mathbb{R}^{2D_{h} \times 1} \)

In this step, a scaling factor \( e \) of size \( L/2 \) is introduced, which is used to scale the dimensions. The result is then processed through the sigmoid activation function to obtain the final output \( V_h \).  The specific formula for this process is as follows:
\[V_h = \sigma \left( \left( V_{l} \cdot W^{L \times e} \right) \cdot W^{e \times L} \right)
\tag{6}\]
Here, \( \sigma \) represents the sigmoid activation function.  \( V_h \) is of size \( L \times 1 \times 1 \), encapsulates the contributions of the different hidden layers in the SSL model, providing an aggregated representation more suitable for the ADD task.

\subsection{Hierarchical Adaptive Mixture of Experts}
Research shows that the mixture of experts (MoE) model can effectively integrate features using a gating network, enhancing the model's performance and flexibility. However, existing MoE models often rely on fixed expert selection strategies. Directly applying them in ADD tasks may cause MoE to overlook the important contributions of multi-level hidden embeddings in the SSL model, especially when dealing with complex patterns and artifacts in spoofing speech.\par
As shown in the Fig.~\ref{fig:HA-MoE}(b), We propose an improved hierarchical adaptive method based on mixture of experts (HA-MoE). The method consists of a learnable hierarchical weight vector $V_h$, a gating network $G$, and an expert network containing $N$ experts $\{E_1, E_2, \dots, E_N\}$. The core improvement is the introduction of a learnable $V_h$ to assist the dynamic routing process of the MoE model, which is used to efficiently integrate the multi-level hidden embedding of the XLSR model and to mine the complex time-frequency patterns and artefacts in the speech spoofing samples.\par
In the HA-MoE method, we perform element-wise multiplication between the hierarchical weight vector \(V_h\) and the hidden embeddings \(x\) from all \(L\) layers using the Hadamard product to obtain the weighted feature \(F\). Then, the feature \(F \in \mathbb{R}^{L \times T \times D}\) is flattened into \(F \in \mathbb{R}^{T \times F}\) and fed into the gating network. Finally, appling a softmax operation to generate the probability distribution for \(N\) experts. The calculation process is as follows:
\[G(x,V_h) = softmax(\delta (\sum_{i=1}^{L}x \circ V_h )\cdot W_G)\tag{7}\]
Here, $\delta$ represents the flatten operation, and $W_G$ is the learnable weight matrix of the gating network, where $W_G \in \mathbb{R}^{F \times N}$.\par
Additionally, the flattened feature \( F \) is also fed into the expert networks \( E_i \). Each expert network consists of two fully connected layers and a ReLU activation layer, which maps \( F \) to the specified output dimensions. Finally, the output of the HA-MoE layer, \( Y(F) \) is the weighted sum of the outputs from the top \( K \) experts, computed as follows:
\[Y(F) = \sum_{i=1}^{K} \frac{TopK(G_i)}{\sum_{j=1}^{K}TopK(G_j)}\cdot E_i(F) \tag{8}\]
where \( G_i \) and \( G_j \) are the weights of the \( i \)-th and \( j \)-th expert, respectively. The \( TopK \) function identifies and retains the highest \( K \) weights, setting the rest to zero. The weights retained by the \( TopK \) function are normalized to ensure their sum equals one.\par
Finally, we use the AASIST classifier to predict the results of audio deepfake detection, and the prediction process is optimized using a weighted cross-entropy loss.

\section{Experimental setup}
\label{sec:pagestyle}
\subsection{Dataset and Evaluation Metrics}
We constructed the AudioFake dataset for the self-supervised learning phase, which is based on high-quality corpus such as LJspeech and VCTK, and generates 810 hours of English spoofed samples using seven commonly used TTS and VC algorithms (including ViTS\cite{ViTs}, HiFi-GAN\cite{Hifigan}, FastDiff\cite{FastDiff}, FreeVC\cite{Freevc}, etc.). All speech samples were resampled to 16 kHz and converted to WAV format, and each speech was randomly trimmed to 4-10 seconds and enhanced by codecs. In addition, we introduced the ASVSpoof5\cite{asvspoof5} challenge dataset, which covers 32 TTS and VC forgery algorithms with additional codecs and compression. The data is aggregated into a total of about 3,000 hours and more than 1.47 million spoofed speech samples, which effectively ensures the diversity and richness of the data.\par
We used the training set of ASVSpoof19LA for fine-tuning, evaluated on four benchmark datasets: ASVSpoofing 19LA\cite{19LA_data}, 21LA, 21DF\cite{21LA_DF_data}, and In-the-wild\cite{Inthewild_data}. The 19LA dataset is a multi-speaker TTS and VC database widely used to assess speech generation verification performance. The 21LA evaluation set contains 181,566 real and synthetic audio samples, with speech affected by dynamic transmission and codec factors. The 21DF evaluation set includes hundreds of generation algorithms not seen in the training set, along with compression variability during audio transmission. The In-the-wild dataset contains real and spoofed speech from over 50 English-speaking celebrities, with similarities in background noise, emotions. Both the 21DF and In-the-wild datasets are better suited for evaluating the generalization ability of ADD systems. We used the minimum tandem detection cost function (min t-DCF)\cite{Min_t-DCF} and the equal error rate (EER) as the primary evaluation metrics for this paper.

\subsection{Implementation Details}
For the SSL stage, we utilized the original pre-training configuration of Wav2vec2.0. The input consisted of full-length speech segments, and we loaded the pre-trained weights of the XLSR-0.3B model to continue training. The learning rate was set to \(1 \times 10^{-5}\), and gradient clipping was applied. Early stopping was triggered if the validation loss failed to decrease for three consecutive epochs. During the fine-tuning phase, all audio inputs were either cropped or padded to a fixed length of 4 seconds. The HA-MoE module was configured with a fixed number of 4 experts, and the Top-k value was set to 2. The AASIST classifier retained its original configuration. We employed the Adam optimizer with \(\beta = [0.9, 0.999]\), an initial learning rate of \(5 \times 10^{-6}\), a weight decay rate of \(1 \times 10^{-4}\), and a batch size of 24. To reduce the class imbalance between real and spoofed samples, we used a weighted cross-entropy loss function and assigned weights of 0.9 and 0.1. In addition, we used Rawboost\cite{Rawboost} for data augmentation. All experiments were performed on 4 NVIDIA A40 GPU.


\section{EXPERIMENTAL}

\begin{table*}[htbp]
\renewcommand{\arraystretch}{1.2}
\centering
\caption{We conducted ablation experiments on the ASVSpoof series and In-the-wild datasets, with results reported in terms of EER (\%). The last column represents the average results across multiple datasets.}
\begin{tabular}{cccccccccc}
\toprule
\multirow{3}{*}{\textbf{Method}} & \multirow{3}{*}{\textbf{Configuration}} & \multirow{3}{*}{\textbf{Trainable Params (M)}} & \multicolumn{4}{c}{\textbf{EER (\%)}} & \multirow{3}{*}{\textbf{EER\textsubscript{Avg} (\%)}} \\
\cmidrule(lr){4-7}
& & & \textbf{ASV-19 LA} & \textbf{ASV-21 LA} & \textbf{ASV-21 DF} & \textbf{In-the-wild} &  \\
\midrule
Full-Param & --            & 317            & 0.323     & 1.647     & 2.673     & 8.584       & 3.307        \\
\midrule
\multicolumn{1}{c}{\multirow{4}{*}{LoRA}} & \multicolumn{1}{c}{r=4}           & \multicolumn{1}{c}{0.66}             & 0.412     & 1.494     & 3.171     & 10.176       & 3.813        \\
\multicolumn{1}{c}{} & \multicolumn{1}{c}{\textbf{r=8}}           & \multicolumn{1}{c}{1.25}             & 0.347     & 1.382     & 2.822     & 9.358       & 3.477        \\
\multicolumn{1}{c}{} & \multicolumn{1}{c}{r=16}          & \multicolumn{1}{c}{2.43}             & 0.295     & 1.327     & 2.967     & 9.384       & 3.493        \\
\multicolumn{1}{c}{} & \multicolumn{1}{c}{r=32}          & \multicolumn{1}{c}{4.79}             & 0.363     & 1.422     & 2.916     & 9.535       & 3.559        \\
\midrule
\multicolumn{1}{c}{\multirow{4}{*}{Adapter}} & \multicolumn{1}{c}{s=16}          & \multicolumn{1}{c}{7.15}             & 0.273     & 1.176     & 2.784     & 8.812       & 3.261        \\
\multicolumn{1}{c}{} & \multicolumn{1}{c}{s=32}          & \multicolumn{1}{c}{14.22}            & 0.198     & 0.952     & 2.727     & 8.476       & 3.088        \\
\multicolumn{1}{c}{} & \multicolumn{1}{c}{\textbf{s=64}}          & \multicolumn{1}{c}{28.43}            & 0.176     & \textbf{0.847}     & 2.481     & 7.883       & 2.847        \\
\multicolumn{1}{c}{} & \multicolumn{1}{c}{s=128}         & \multicolumn{1}{c}{56.69}            & 0.212     & 0.872     & 2.563     & 8.163       & 2.953        \\
\midrule
Hybrid     & r=8, s=64     & 29.68            & \textbf{0.167}     & 0.943     & \textbf{2.431}     & \textbf{7.762}       &  \textbf{2.826}        \\
\bottomrule
\end{tabular}
\label{tab:ssl_experiment}
\end{table*}

\subsection{Main Results}
Table~\ref{tab:ssl_experiment} compares the performance of different self-supervised pretraining methods. All methods use full-tuning in the fine-tuning stage. Experiments based on the LoRA method show that when $r = 8$, the model achieves an average EER of 3.477\% with only 1.25M trainable parameters. However, due to the small number of parameters, the model struggles to capture complex forged features, resulting in an overall performance that is not as good as that of full pretraining. For the Adapter method, when $s = 64$, the trainable parameters are increased to 28.43M, and the average EER decreases to 2.847\%, which outperforms the full-parameter pretraining. This indicates that the introduction of the adapter method preserves the original pretraining knowledge and avoids catastrophic forgetting. However, increasing the dimensionality further to $s = 128$ leads to performance degradation, suggesting that too high a dimensionality may introduce more noise and redundant information, which reduces the model's generalization ability. Finally, by combining the LoRA and Adapter methods, we achieved the best results with an average EER of 2.826\%. The hybrid method better captures global and local spoofing cues and effectively improves the generalization ability of the self-supervised features. This is especially evident on the more challenging ASV-21 DF and In-the-wild datasets. Furthermore, with 29.68M trainable parameters, the hybrid method strikes a good balance between performance and training efficiency.

\subsection{Comparison with the State-of-the-art Systems}
In Table~\ref{tab:asv_comparison}, we compare the performance of the proposed Wav2DF-TSL system with other state-of-the-art methods on ASVSpoof 2021 LA and DF datasets.  The results show that Wav2DF-TSL achieves EER of 0.87\% and 1.95\% on the 21LA and 21DF datasets, respectively, outperforming all existing systems.
In addition,  all methods are based on self-supervised features, our method integrates adapter modules into the XLSR model, demonstrating that the self-supervised learning process with unlabeled spoofing data can effectively enhance the representation of front-end features. Compared to the system using the AASIST classifier, the performance of Wav2DF-TSL shows a significant improvement. On the more challenging 21DF dataset, the EER is reduced from 2.87\% to 1.95\%, achieving a 31\% relative improvement, validating the effectiveness of our method.\par
In Table~\ref{tab:in_the_wild}, we compare the proposed Wav2DF-TSL system with other state-of-the-art methods on the In-the-wild dataset. Due to the presence of background noise and similar sentiment cues in the data, the model requires stronger out-of-domain generalization capabilities. The results show that Wav2DF-TSL achieves an EER of 6.83\%, demonstrating optimal performance and validating the model's generalization and robustness across various spoofing scenarios. Compared to the XLSR-SLS method, which also uses multi-layer feature fusion, the EER is reduced from 8.87\% to 6.83\%, achieving a 23\% relative improvement. The HA-MoE method we introduced fuses multi-layer features through multiple expert routing, further capturing the multi-level spoofing cues. Notably, this is SOTA performance achieved on the In-the-wild dataset.

\begin{table}[h!]
\centering
\caption{Comparison with other ADD systems on the ASVSpoof2021 LA and DF eval set, reported in terms of min t-DCF and EER(\%).}
\fontsize{8}{12}\selectfont  
\renewcommand{\arraystretch}{1}  
\setlength{\parskip}{0pt}  
    \centering
\begin{tabular}{lccc}
\hline
\multirow{2}{*}{\textbf{System}} & \multicolumn{2}{c}{\textbf{ASV-21 LA}} & \textbf{ASV-21 DF} \\ 
\cline{2-4}
                                 & \textbf{min t-DCF} & \textbf{EER(\%)}       & \textbf{EER(\%)}       \\ \hline
XLSR+LLGF\cite{XLSR+LLGF}                        & --               & 7.62                 & 5.44                 \\
XLSR+AASIST\cite{XLSR+AASIST}                      & 0.2121           & 0.98                 & 2.87                 \\
XLSR+AASIST2\cite{XLSR+AASIST2}                     & --               & 1.61                 & 2.77                 \\
XLSR+Conformer\cite{XLSR+conformer}                   & 0.2116           & 0.97                 & 2.58                 \\
WavLM+MFA\cite{WavLM+MFA}                        & --               & 5.08                 & 2.56                 \\
XLSR+ACS\cite{XLSR+ACS}                         & 0.2172           & 1.30                 & 2.19                 \\
XLSR+Conformer+TCM\cite{XLSR+TCM}               & 0.2130   & 1.03      & 2.06        \\
\hline
\textbf{Wav2DF-TSL(ours)}       & \textbf{0.2082}  & \textbf{0.87}        & \textbf{1.95}        \\ \hline
\end{tabular}
\label{tab:asv_comparison}
\end{table}

\begin{table}[h!]
\centering
\caption{Comparison with other ADD systems on the In-the-wild evaluation set, reported in terms of EER(\%).}
\fontsize{8}{13}\selectfont  
\renewcommand{\arraystretch}{1}  
\setlength{\parskip}{0pt}  
\begin{tabular}{p{0.33\textwidth}c}
\hline
\textbf{System}         & \textbf{In-the-wild} \\
\hline
XLSR,WavLM,Hubert+ResNet18\cite{XLSR_WavLM_Hubert+Fusion}         & 24.27               \\
XLSR+Linear\cite{XLSR+Linear}                 & 16.17               \\
XLSR+Voc\cite{XLSR+Voc}                & 12.32               \\
XLSR+SLS\cite{XLSR+SLS}               & 8.87                \\
OCKD\cite{OCKD}                    & 7.68                \\
\hline
\textbf{Wav2DF-TSL(ours)}       & \textbf{6.83}       \\ 
\hline
\end{tabular}
\label{tab:in_the_wild}
\end{table}

\begin{table*}[htbp]
\fontsize{8}{12}\selectfont  
\renewcommand{\arraystretch}{1}  
\setlength{\parskip}{0pt}  
\centering
\caption{We conducted ablation experiments on the ASVSpoof series and In-the-wild datasets, with results reported in terms of EER (\%). Here, FT SSL-Model refers to the range of weight updates for the SSL model during the fine-tuning stage, and the last column represents the average results across multiple datasets.}
\begin{tabular}{cccccccccc}
\bottomrule
\multirow{3}{*}{\textbf{ID}} & \multirow{3}{*}{\textbf{SSL-Hybrid}} & \multirow{3}{*}{\textbf{HA-MoE}} & \multirow{3}{*}{\textbf{FT SSL-Model}} & \multirow{3}{*}{\textbf{Params(M)}} & \multicolumn{4}{c}{\textbf{EER (\%)}} & \multirow{3}{*}{\textbf{EER\textsubscript{Avg} (\%)}} \\
\cmidrule(lr){6-9}
 &  &  &  &  & \textbf{ASV-19 LA} & \textbf{ASV-21 LA} & \textbf{ASV-21 DF} & \textbf{In-the-wild} &  \\
\bottomrule
A1 & $\times$ & $\times$ & Full-Turning          & 317.8 & 0.241 & 0.983 & 2.873 & 9.416 & 3.378 \\
A2 & × & \checkmark & Full-Turning          & 325.9 & 0.193 & 0.844 & 2.612 & 8.476 & 3.031 \\
A3 & \checkmark & × & Full-Turning          & 347.3 & 0.167 & 0.943 & 2.431 & 7.762 & 2.826 \\
A4 & \checkmark & \checkmark & Full-Turning          & 355.4 & \textbf{0.103} & 0.872 & \textbf{1.954} & \textbf{6.827} & \textbf{2.439} \\
\midrule
B1 & \checkmark & \checkmark & Frozen SSL            & 8.5   & 0.362     & 1.214     & 3.157     & 10.652     & 3.846     \\
B2 & \checkmark & \checkmark & Adapter Layer         & 38.2  & 0.126 & \textbf{0.782} & 2.321 & 7.643 & 2.718 \\
\bottomrule
\label{tab:Ablation_study}
\end{tabular}
\end{table*}


\begin{figure}[htbp]
  \centering
  \begin{subfigure}{0.49\columnwidth}
    \includegraphics[width=\columnwidth]{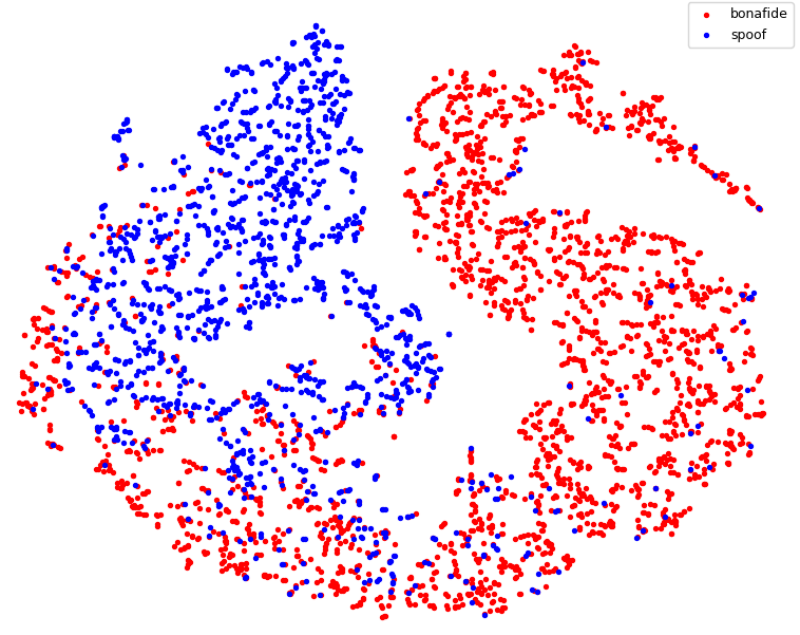}
    \caption{Baseline}
    \label{fig:sub1}
  \end{subfigure}
  \hfill
  \begin{subfigure}{0.49\columnwidth}
    \includegraphics[width=\columnwidth]{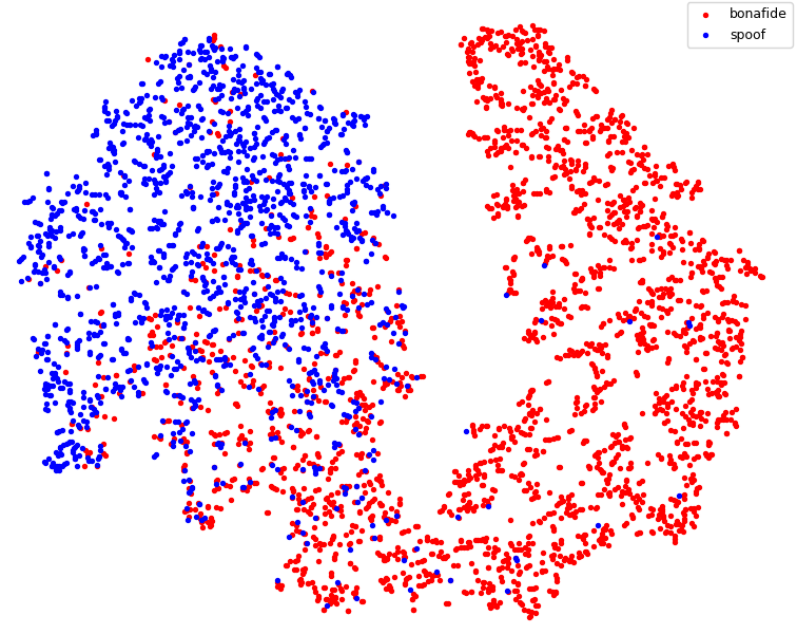}
    \caption{SSL-Hybrid}
    \label{fig:sub2}
  \end{subfigure}

  \vspace{0.1cm}

  \begin{subfigure}{0.49\columnwidth}
    \includegraphics[width=\columnwidth]{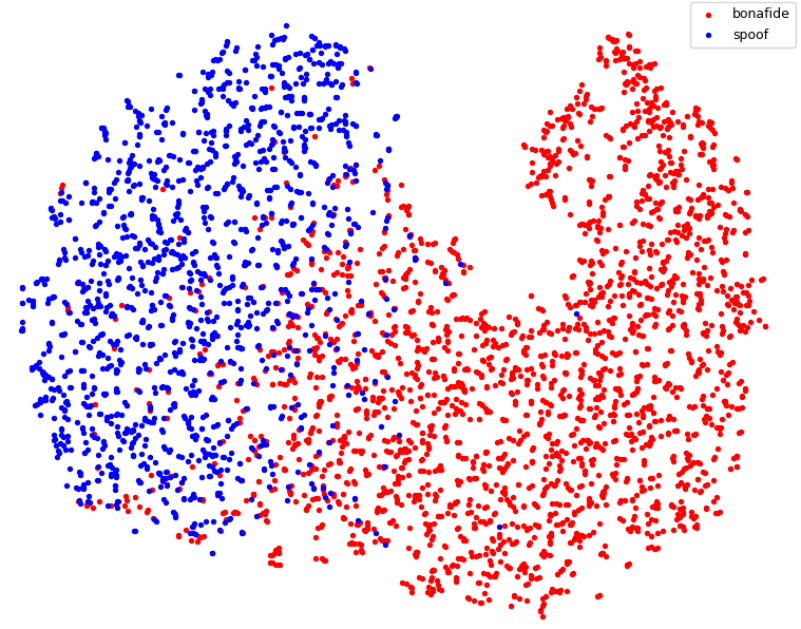}
    \caption{HA-MoE}
    \label{fig:sub3}
  \end{subfigure}
  \hfill
  \begin{subfigure}{0.49\columnwidth}
    \includegraphics[width=\columnwidth]{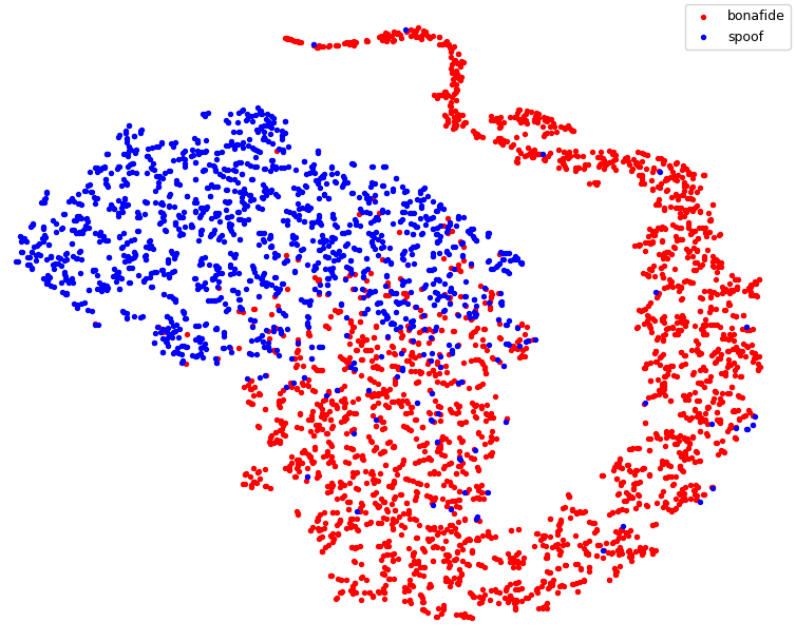}
    \caption{Wav2DF-TSL}
    \label{fig:sub4}
  \end{subfigure}

  \caption{Visualization of t-SNE embeddings from the In-the-wild dataset compares the performance of the proposed system. Red represents real speech (bonafide) and blue represents spoofed speech.}
  \label{fig:T-SNE}
\end{figure}

\subsection{Ablation Study}
The ablation experiments validate the effectiveness of the proposed two-stage learning strategy, as shown in Table~\ref{tab:Ablation_study}. Experiment A1 presents the baseline results without any enhancements. Experiment A2 applies the HA-MoE method alone, achieving an average EER reduction of 10.3\% compared to the baseline. This demonstrates that by integrating multi-level self-supervised hidden features, the model can capture spoofing cues at different levels, thereby improving its ability to distinguish between bona fide and spoofed speech. Experiment A3 demonstrates that after introducing an adapter-based efficient self-supervised pre-training strategy, the model achieved an average EER of 2.826\%, representing a relative improvement of 16.3\% compared to the baseline. This indicates that incorporating task-specific synthetic data into the self-supervised learning process effectively enhances the SSL model's ability to represent spoof-related features. Furthermore, it improves the model's generalization and robustness, particularly in the more challenging datasets such as ASV-21 DF and In-the-wild. Experiment A4 demonstrates that combining the two-stage learning strategy, achieving an average EER of 2.439\%. Compared to the baseline, it achieves a relative improvement of 27.8\% on average, validating the compatibility and effectiveness of the proposed method. Additionally, as shown in Figure~\ref{fig:T-SNE}(d), Wav2DF-TSL produces a more compact and clearer decision boundary compared to the baseline and single-stage methods.
Experiment B1 shows that with the SSL model frozen during fine-tuning, the learnable parameters are only 8.5 M. However, too low a number of parameters limits the expressive power of the model, leading to a significant degradation of its performance compared to the baseline on multiple evaluation sets. Experiment B2 shows that updating only the adapter layer during fine-tuning reduces the average EER to 2.718\%, a 19.5\% decrease compared to the baseline, while using only 12\% of the parameters. The performance improvement mainly stems from the efficient update of the adapter layer rather than an increase in parameter count, highlighting its critical role in performance optimization.
\section{CONCLUsION}

In this paper, we proposed a novel two-stage learning strategy based on the XLSR pre-trained model, designed for the task of audio deepfake detection, called Wav2DF-TSL. In the self-supervised learning stage, we integrate LoRA and adapters while freezing other parameters, optimizing the SSL model's ability to adapt to spoofed features and artifacts while mitigating catastrophic forgetting. In the fine-tuning stage, We proposed the HA-MoE method to process multi-layer hidden embeddings by enhancing the gating network's focus on hierarchical information and dynamically selecting the most suitable experts for feature fusion, effectively improving the model's task adaptability. Experiments show that the proposed method consistently improves performance on four benchmark datasets. Using only 12\% of the learnable parameters, it achieves a 19.5\% relative reduction in average EER, demonstrating its ability to effectively enhance training efficiency while maintaining strong generalization and robustness. In future work, we will explore model distillation methods to further optimize the system's inference efficiency.

\bibliographystyle{ieeetr}  
\bibliography{IEEEabrv, references}  

\end{document}